\documentclass[twocolumn,english,prl,aps,superscriptaddress,amsmath,amssymb,floatfix]{revtex4-1}
\usepackage[T1]{fontenc}
\usepackage[latin9]{inputenc}
\usepackage{color}
\usepackage{babel}
\usepackage{amssymb}
\usepackage{graphicx}
\usepackage{subscript}
\usepackage[unicode=true,pdfusetitle,
 bookmarks=true,bookmarksnumbered=false,bookmarksopen=false,
 breaklinks=false,pdfborder={0 0 0},pdfborderstyle={},backref=false,colorlinks=true]
 {hyperref}
\hypersetup{
 colorlinks,linkcolor=red,citecolor=blue}
\begin{document}
\title{Experimental demonstration of multimode microresonator sensing by
machine learning}
\author{Jin Lu}
\address{College of Information Engineering, Zhejiang University of Technology,
Hangzhou 310023, P. R. China.}
\author{Rui Niu}
\affiliation{CAS Key Laboratory of Quantum Information, University of Science and
Technology of China, Hefei 230026, P. R. China.}
\author{Shuai Wan}
\affiliation{CAS Key Laboratory of Quantum Information, University of Science and
Technology of China, Hefei 230026, P. R. China.}
\author{Chun-Hua Dong}
\affiliation{CAS Key Laboratory of Quantum Information, University of Science and
Technology of China, Hefei 230026, P. R. China.}
\author{Zichun Le}
\address{College of Science, Zhejiang University of Technology, Hangzhou 310023,
P. R. China.}
\author{Yali Qin}
\address{College of Information Engineering, Zhejiang University of Technology,
Hangzhou 310023, P. R. China.}
\author{Yingtian Hu}
\address{College of Information Engineering, Zhejiang University of Technology,
Hangzhou 310023, P. R. China.}
\author{Weisheng Hu}
\address{State Key Laboratory of Advanced Optical Communication Systems and
Networks, Shanghai Jiao Tong University, Shanghai 200240, P. R. China.}
\author{Chang-Ling Zou}
\email{clzou321@ustc.edu.cn}

\affiliation{CAS Key Laboratory of Quantum Information, University of Science and
Technology of China, Hefei 230026, P. R. China.}
\author{Hongliang Ren}
\email{hlren@zjut.edu.cn}

\address{College of Information Engineering, Zhejiang University of Technology,
Hangzhou 310023, P. R. China.}
\date{\today}
\begin{abstract}
\textcolor{black}{A multimode microcavity sensor based on a self-interference
microring resonator is demonstrated experimentally. The proposed multimode
sensing method is implemented by recording wide band transmission
spectra that consist of multiple resonant modes. It is different from
}{previous dissipative sensing }\textcolor{black}{scheme, which}{ 
aims at measuring the transmission depth changes of}\textcolor{black}{ 
a single resonant mode in a microcavity.}{  }\textcolor{black}{Here,
by combining the dissipative sensing mechanism and the machine learning
algorithm,}{  the multimode sensing information extracted from }\textcolor{black}{a
broadband spectrum}{  can be efficiently fused to estimate the target
parameter}\textcolor{black}{.}{  The multimode sensing method is
}\textcolor{black}{immune to laser frequency noises and robust against
system imperfection,} thus our work presents a great step towards
practical applications of microcavity sensors outside the research
laboratory.\textcolor{black}{  The voltage applied across the microheater
on the chip was adjusted to bring its influence on transmittance through
the thermo-optic effects. As a proof-of-principle experiment, the
voltage was detected by the multimode sensing approach. The experimental
results demonstrate that the limit of detection of the multimode sensing
by the general regression neural network is reduced to 6.7\% of that
of single-mode sensing within a large measuring range.} 
\end{abstract}
\maketitle

\section{INTRODUCTION}

Optical microphotonic sensors have excellent performance in terms
of high responsivity, high sensitivity, and small foot-print{$\,$\citep{key-9-8,2-1,1}}.
One specific class of the microphotonic sensors, the so-called microcavity
sensor supporting whispering gallery modes (WGMs), has attracted much
research interest in recent years$\,$\citep{key-1,key-11,2,key-5-18-1,key-5-18-2}.
With ultrahigh optical quality (Q) factors and ultrasmall mode volume,
strongly localized WGMs lead to significant enhancement of light-matter
interaction. So the environment perturbations, such as pressure, temperature,
and force, could be detected with great sensitivity in WGM sensors.
Typical microcavity sensing is implemented by measuring the resonant
spectral changes, including the induced changes in resonant frequency,
spectrum linewidth, and transmission depth~\citep{3-1,3,key-1-11-1,K-4}.
In conventional waveguide-coupled WGM sensors, the reactive interaction
could only induce a frequency shift, which could be detected by high
precision scanning lasers. Due to the limited scanning range and slow
scanning rate, only a single optical resonance was measured to perform
the sensing in most previous studies~\citep{K-4}.

However, the single-mode sensing performance might be restricted with
the single information channel~\citep{3-1,3}. In a WGM microcavity,
various WGMs in wavelength are present with different azimuthal mode
numbers, polarizations, quality factors, and so on. Generally speaking,
these WGMs provide diverse sensing modalities. They are just like
different sensing information collected by multiple sensors in the
another sensing scene, where a target parameter is also detected by
multiple sensor sources. By the aid of the existing sensor fusion
technology, these different informations from multiple sensors could
be effectively merged so that the resulting target parameter has less
uncertainty than would be possible when these sensors were used individually~\citep{key-31}.
Therefore, the sensing performance can be improved significantly by
employing multiple WGMs, since the information that associate with
other high-Q resonances are collected~\citep{key-1-1,key-1-2,key-1-3,key-1-10,key-1-11}.
In practice, there are two challenges to preventing such multimode
sensing: (1) As noted above, scanning a wide range spectrum is a time-consuming
process and also demanding a high-cost scanning laser. Besides, the
laser frequency may be not stable and the cavity frequencies are also
sensitive to thermal fluctuations~\citep{key-1-2}. (2) The responses
of different WGMs to an unknown target parameter are different in
general. This is due to the material and geometry dispersion of the
microcavity WGMs, and also wavelength-dependent light-matter interaction.
Without the sensing fusion algorithm, it is also difficult to estimate
a parameter by measuring the responses of different WGMs. Therefore,
the analysis of the full spectrum is regarded as time- and resource-consuming.

In this work, self-interference microring resonator (SIMRR) based
multimode sensing is experimentally demonstrated. With the help of
the dissipative sensing mechanism~\citep{key-1-6,key-5-18-4}, a
machine learning algorithm is used to fuse the multimode sensing data
of a SIMRR. Because dissipative sensing is robust against frequencies
noises, transmission depths of multiple resonant modes or all the
intensity samples in the spectra, instead of their resonant frequencies,
are considered as the sensing data. The SIMRR based multimode sensing
principle has been theoretically proposed in Ref.~\citep{key-1-5}.
By probing the experimental system with a broadband laser source,
these transmission depths of multiple resonant modes or full spectra
are easily collected for training the artificial neural network (ANN)
and estimating the object parameter. The trained ANN develops a prediction
model. After the model is validated with the test set, its output
is the ``estimated'' measurement result. It is worth noting that
the spectra are not fitted in all data processing. Compared with the
single-mode dissipative sensing, the SIMRR-based multimode sensing
method improves the detecting accuracy considerably. The proposed
method has great potential to offer a simple, robust, and high-sensitive
sensing platform with low detecting cost.

\section{MUlTIMODE SENSING MECHANISM }

The conventional add-drop microring resonator (ADMRR) consists of
a microring resonator sandwiched between two parallel waveguides.
In the typical amplitude transmittance of an ADMRR, a Lorentzian line
shape is periodically repeated along the wavelength axis. Owing to
almost the same spectral shifts in multiple resonances, it is often
used to monitor the target parameter by measuring a resonant frequency
shift, which is referred to as dispersive sensing. As demonstrated
in our previous works~\citep{key-1-6,key-5-18-4,key-1-5,key-14-1},
the SIMRR's spectrum shows a feature of modulated extinction ratio
over a large wavelength range, and the SIMRR is more robust against
the laser frequency noises by converting frequency shift into changing
of extinction ratio. The dissipative sensing based on SIMRR is enhanced
by the Vernier effect due to the interference effect.

\begin{figure}[tp]
\centerline{\includegraphics[width=1\columnwidth]{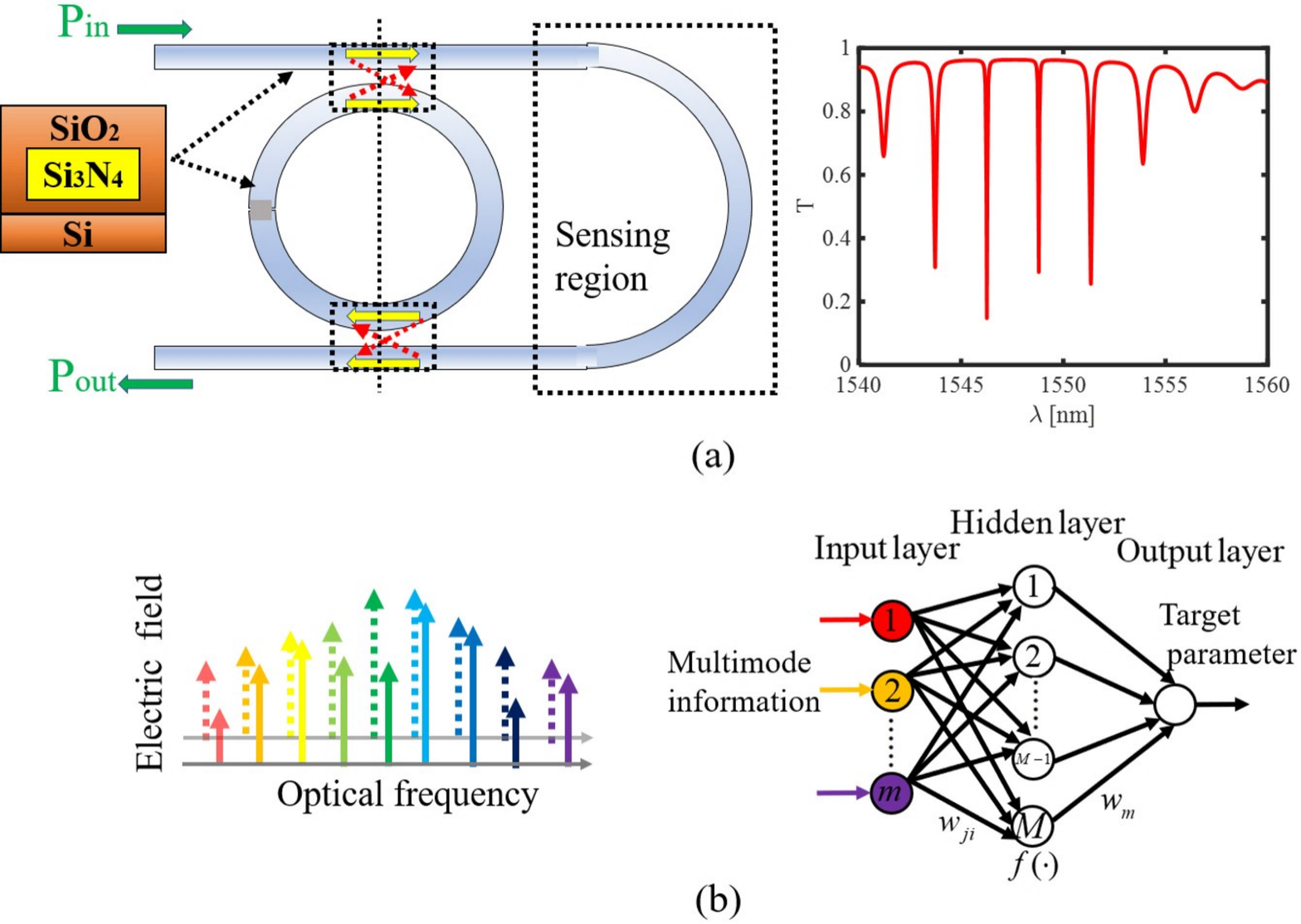}}\caption{The proposed multimode sensing mechanism\textcolor{black}{. (a) }Structural
diagram of a SIMRR-based sensor design and typical amplitude transmittance
of a SIMRR. (b) An ANN assisted multimode sensing information fusion
for predicting the target parameter.}
\label{Fig.1}
\end{figure}

Therefore, the SIMRR-based multimode sensor is taken as an example,
and the principle of the proposed multimode microcavity sensor is
explained. As shown in Fig.\ref{Fig.1}(a), the SIMRR consists of
is a microring resonator that is coupled by a sensing arm waveguide
two times. When a SIMRR is used as a sensor, the sensing arm waveguide
is exposed to the target analyte to be detected. The sensing target
can trigger the phase shift with different wavelengths under an optical
path of the sensing arm waveguide. Eventually, distinct phenomena
are exhibited in its transmission spectra, including the modulated
transmission depth and spectrum linewidth. Its typical transmittance
is displayed in the inset of Fig.\ref{Fig.1}(a). In contrast to the
ADMRR, the SIMRR shows a strong wavelength-dependent response~\citep{key-41,key-51}.
Although optical modes with different mode family, polarization, and
orbital angular momentum can be excited in a WGM microcavity, the
multimode sensing method is mainly hindered by the difficulty of full-spectrum
analysis employing a frequency tracking method. Consequently, the
existing WGM sensing methods, including dispersive and dissipative
sensing, always focus on measuring the spectral change at a single
resonance and results in the throw away of spectral shift data analysis
in other resonances.

Based on previous efforts on the single-mode dissipative sensing,
the multimode sensing has been theoretically investigated in SIMRR-based
sensor~\citep{key-1-5}. Theoretical studies predict a significant
enhancement of the detection accuracy. The single-mode sensing can
be performed with different sensitivity based on the spectral change
of each resonance. These sensitivities have to be estimated and accounted
for in the process of estimating an eventual result. Therefore, one
key issue is how to merge the sensing information from multiple resonances.
In the sensor fusion technology, the neural network is particularly
well suited for the combination of multiple sensing information~\citep{key-31,key-5-18-5,key-61,key-5-18-3,key-33}.
Hence, we propose the supervised machine learning algorithm to fuse
these multimode sensing informations. As shown in Fig.\ref{Fig.1}(b),
a back propagating-ANN (BP-ANN) is used to merge these transmission
depth changes from multiple resonances. The multimode fusion model
based on BP-ANN is built to estimate the target parameter. In the
input layer, the input variables are more accurately described as
an input set, since it generally consists of multiple independent
variables (multiple transmission depth changes from multiple resonances),
rather than a single value. A hidden layer is located between the
input and output layers of the algorithm, where the function applies
weights $w_{ji}$ to the inputs and directs them through an activation
function $f(\cdot)$ as its output. In the output layer, the ultimate
output is the summation of the product of the output signals in the
hidden layer and another weight factor ($w_{m}$), and it is the estimated
value of the target parameter.

The detailed process of implementation is as follows. Firstly, the
related original data, including training and test sets, are collected
for the BP-ANN. To train the network, the supervised learning algorithm
requires a training set of paired inputs and known output labels.
Then, the training set needs to be collected in advance. That is to
say, many groups of transmission depths of multiple resonances are
experimentally collected with known target parameter values before
the measurement is started. After this, the actual measurement is
performed by the SIMRR-based sensor, and an additional group of transmission
depths is collected as the test set. Secondly, the network is trained
by a back-propagation algorithm, which adjusts weights according to
the gradient descent method~\citep{key-33}. The training process
is completed after the training goal error is achieved. The purpose
of the training is to establish the nonlinear mapping relationship
between the input variables (a group of transmission depths) and the
output variable (corresponding target parameter). Finally, the generalization
ability of the trained network is investigated. The test set is input
into the trained network, and the eventual measurement result is the
output by BP-ANN. Here, it is emphasized that the training set should
be collected before the measurement is performed by the SIMRR-based
sensor in practice, so the real-time performance may be preserved.

The proposed multimode sensing method aided by the machine learning
algorithm can be exploited in other multimode microcavity resonators.
The multimode sensing information can be selected as the changes in
resonant frequency, transmission depth, and spectrum linewidth of
multiple resonances or their combination. The multimode sensing information
fusion technology can also be performed with the help of the other
effective fusion algorithms~\citep{key-5-18-3}. In general, the
proposed multimode sensing can achieve higher accurate results compared
to the single-mode sensing.

\section{DEVICE FABRICATION AND EXPERIMENTAL SETUP }

\begin{figure}[tp]
\centerline{\includegraphics[width=1\columnwidth]{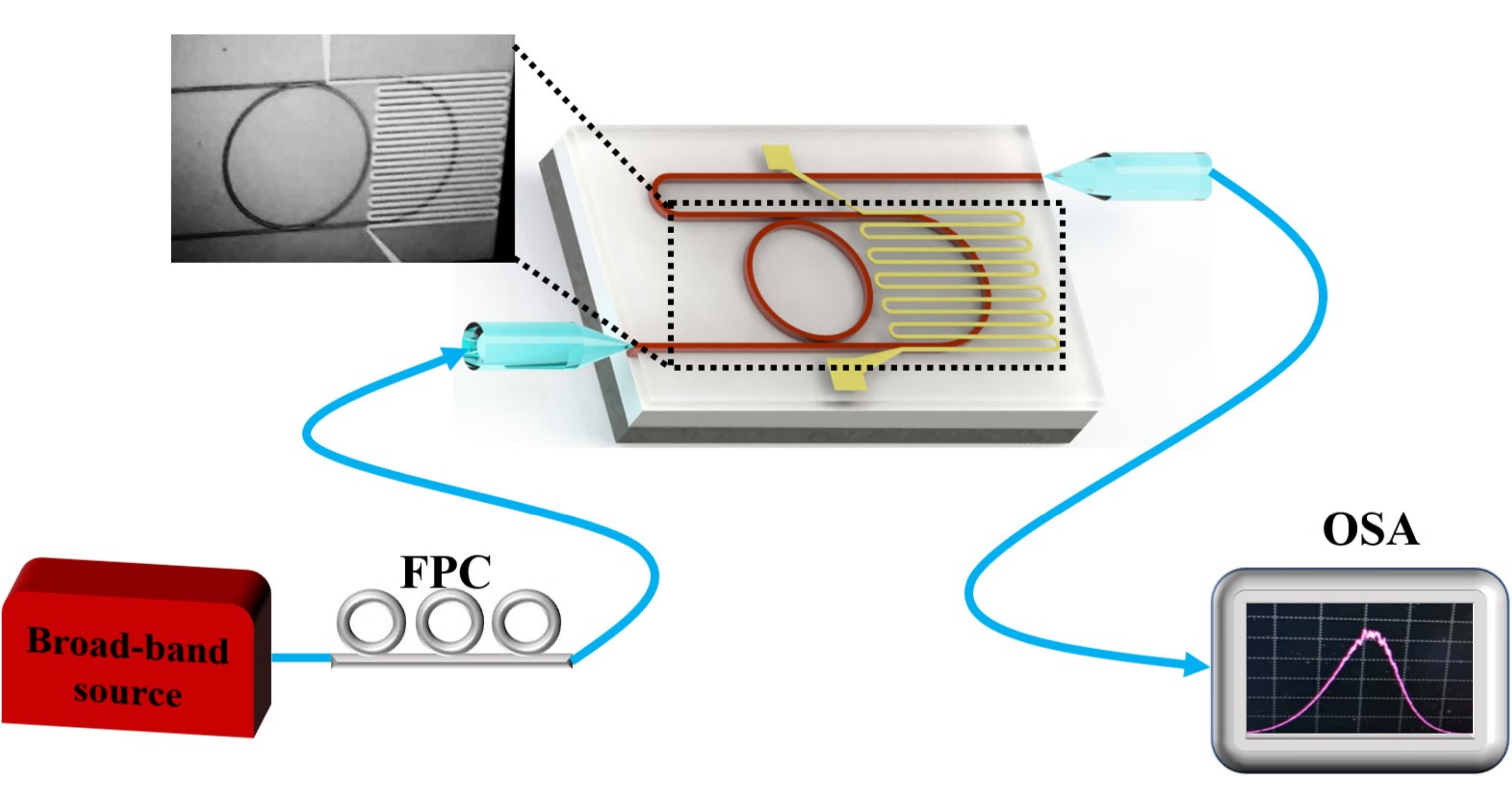}}\caption{Schematic of the experimental setup of self-interference microring
sensor launched by a broad-band laser source. FPC, fiber polarization
controller, OSA, optical spectrum analyzer. An optical microscopy
picture of the SIMRR device is shown in the inset of the figure, and
the microheater is twisted around the upper cladding right above the
sensing arm waveguide.}
\label{Fig.2}
\end{figure}

To demonstrate the multimode sensing, we fabricated a SIMRR sensor
based on a Si\textsubscript{{\footnotesize{} 3}}N\textsubscript{{\footnotesize{} 4}}
wafer. The inset of Fig.\ref{Fig.2} displayed an optical microscopy
picture of the fabricated SIMRR sensor, with a microheater composed
of a metal stripe that zigzagged its way across the upper cladding
right above the sensing waveguide arm. The wafer consists of a bottom
layer of $500\,\mathrm{\mu m}$ silicon, a middle layer of $3\,\mathrm{\mu m}$
wet oxidation SiO\textsubscript{{\footnotesize{} 2}}, and a top layer
of $350\,\mathrm{nm}$ Si\textsubscript{{\footnotesize{} 3}}N\textsubscript{{\footnotesize{} 4}}.
E-beam lithography with hydrogen-silsesquioxane (HSQ) resist was used
to write the circuit pattern of SIMRR into the top layer of the wafer.
Then an inductively-coupled plasma (ICP) source was used to etch Si\textsubscript{{\footnotesize{} 3}}N\textsubscript{{\footnotesize{} 4}}
films with O\textsubscript{{\footnotesize{} 2}}/CHF\textsubscript{{\footnotesize{} 3}}
gas mixtures. After stripping the resist with a buffered oxide etchant
(BOE), a $3\,\mathrm{\mu m}$ thick silicon dioxide layer, acting
as an upper cladding, was deposited on the Si\textsubscript{{\footnotesize{} 3}}N\textsubscript{{\footnotesize{} 4}}
films by a plasma-enhanced chemical vapor deposition (PECVD) process.
As shown in the inset of ~Fig.\ref{Fig.1}(a), the cross-section
of the microring and waveguide was $2\,\mathrm{\mu m}\times350\,\mathrm{nm}$.
By using e-beam evaporation, a metal layer of $300\,\mathrm{nm}$
Au and $10\,\mathrm{nm}$ Ti was successively deposited and twisted
around the upper cladding right above the U-shaped sensing arm waveguide
to form the microheater. The radius of the microring was $100\mathrm{\,\mu m}$,
and the initial length of the sensing arm waveguide was $250\,\mathrm{\mu m}$.
The gap between the microring and waveguide was $200\,\mathrm{nm}$,
and the waveguide loss coefficient $\alpha$ was about $0.67\,\mathrm{dB/cm}$.
Here, the voltage applied across the microheater is directly selected
as the target parameter. It is noted that the temperature of the buried
sensing arm waveguide was not calibrated in this work because it was
difficult to directly measure its temperature.

The experimental setup for SIMRR-based multimode sensing is shown
in Fig.\ref{Fig.2}. The detecting system was excited by a broad-band
laser light source (CLD 1015, Thorlabs) with over $120\,\mathrm{nm}$
spectral bandwidth. Two tapered lens fiber components were used to
couple the power into and extract the power out of the SIMRR, respectively.
The measured coupling loss of lens fibers is $3\,\mathrm{dB/facet}$.
Compared with transverse magnetic (TM) {polarization WGM mode, the
}transverse {electric (TE) mode has a higher quality factor and
sensitivity. In the experiment, a polarization controller located
before the }first tapered lens fiber{ was used to adjust the polarization
state of the cavity mode. The TE mode was selected by comparing the
quality factor of different polarization}\textcolor{black}{. }An
optical spectrum analyzer (OSA) was used to acquire the spectra containing
multiple resonances. By adjusting the applied voltage within the set
range, many groups of transmission depths were extracted as the original
data for the multimode sensing. \textcolor{black}{It is expected that
the cost of multimode sensor would be further reduced in the future,
by employing a charged coupled device (CCD) and a reflection grating
to directly acquire the cavity mode intensities~\citep{key-71},
instead of using the OSA for measuring the transmission spectra~\citep{key-15,key-16}.}

\section{EXPERIMENTAL RESULTS AND MULTIMODE SENSING BASED ON ARTIFICIAL NEURAL
NETWORK }

\subsection{Raw data collection}

\begin{figure*}[tp]
\centerline{\includegraphics[width=1.5\columnwidth]{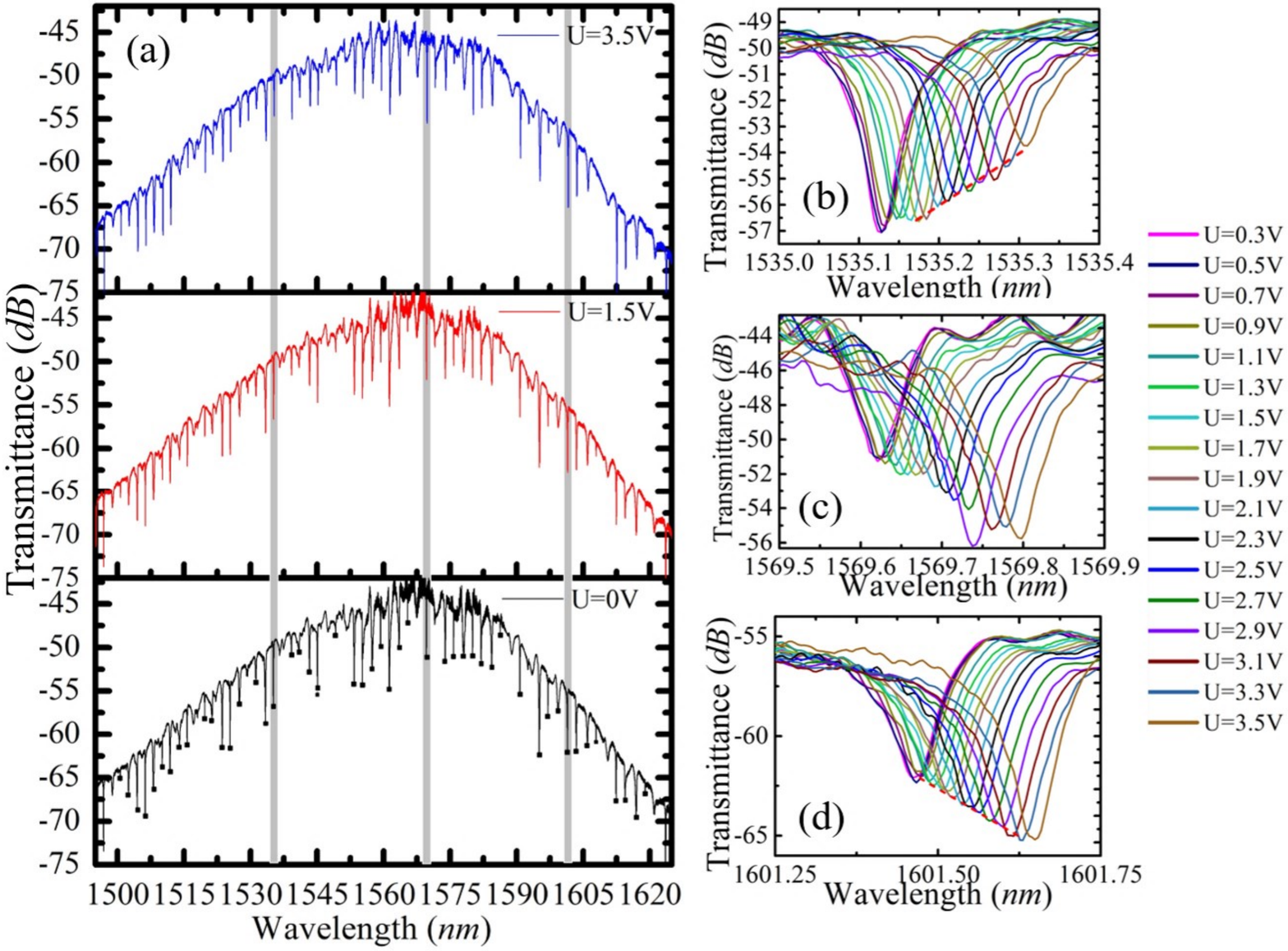}}\caption{(a) With the micro-heater driven by various DC voltages (U=0, 1.5,
3.5$\,$\textit{V}), the experimentally measured transmission spectra
of the SIMRR sensor excited by a broadband laser source, and the transmission
dips near $1535.2\,\mathrm{nm}$, $1569.7\,\mathrm{nm}$ and $1601.5\,\mathrm{nm}$
are indicated by the shadows. When the DC voltages applied across
the micro-heater is changed from 0.3$\,$V to 3.5$\,$V with the step
size of 0.2$\,$V, the experimental transmission spectra near $1535.2\,\mathrm{nm}$
(b), $1569.7\,\mathrm{nm}$ (c), and $1601.5\,\mathrm{nm}$ (d).}
\label{Fig.3}
\end{figure*}

Since the SIMRR shows a strong wavelength-dependent response, its
broadband spectra contain rich frequency domain information. These
transmission depths of multiple resonances extracted from the spectra
were considered as the original data. They were processed by a supervised
machine learning algorithm as described above. First of all, the processing
models built by the BP-ANN required sufficient labeled data for training~\citep{key-2-11,key-17}.
To do so, we varied the DC voltages applied across the microheater
and collected the corresponding transmission spectra of the SIMRR
sensor. Here, the voltage generated by a DC source could be tuned
from 0 to $4\,\mathrm{V}$ with a precision of $0.1\,\mathrm{V}$.
Fig.\ref{Fig.3}(a) shows the measured transmission spectra for the
cases of the voltage applied across the micro-heater U=0, 1.5, 3.5$\,$V.
Here, only transmission spectra at three applied voltages are displayed
so that the spectral variations are easy to be identified in the stacked
figures. Within the wavelength range from $1500\,\mathrm{nm}$ to
$1620\mathrm{\,nm}$, there are multiple strong and narrow resonant
dips present. These dips represent resonant modes inside the SIMRR
excited by the broad-band laser source. Transmission spectra at U=0$\,$V
are taken as an example to illustrate these resonant modes, which
are marked by the square labels at the corresponding transmission
depths. In the wavelength range, the spectra containing \textcolor{black}{49
r}esonant modes need to be collected for the sensing purpose. In the
experiment, a variable optical attenuator was used to control the
power of the laser at the level of $100\,\mathrm{\mu W}$. Then, with
such low excitation power, the laser was coupled into the SIMRR through
tapered lens fiber. The power level caused that the nonlinear effect
in the microcavity could be neglected. Moreover, the frequency interval
between these modes is enough large, and the interactions between
these modes cannot occur. Besides, these transmission dips in transmission
spectra were exhibited based on a plurality of different baselines.
The baselines were formed mainly due to the wavelength-dependent output
intensity distribution of the broad-band laser source. The other part
is due to the wavelength sensitive fiber coupling efficiency in the
detecting system.

Fig.\ref{Fig.3}(b)-(d) separately displayed the transmission dips
near $1535.2\,\mathrm{nm}$,$1569.7\,\mathrm{nm}$ and $1601.5\,\mathrm{nm}$
when the DC voltage applied across the microheater is changed from
0.3$\,$V to 3.5$\,$V with the step size of 0.2$\,$V. They were
indicated by the attached shadows in Fig.\ref{Fig.3}(a). As shown
in Fig.\ref{Fig.3}(b), for the transmission dip near $1535.2\,\mathrm{nm}$,
its transmission depth gets bigger with the increasing of the applied
DC voltage. In Fig.\ref{Fig.3}(c), the transmission depth near $1569.7\,\mathrm{nm}$
does not show a monotonous change with the change of the applied DC
voltage. When the applied voltages were larger than 2.9$\,$$\mathrm{V}$,
the baselines showed a significant reduction in their intensities
away from resonance. The transmission intensity fluctuation maybe
stems from the interference instability inside the SIMRR. Therefore,
it is unable to use the transmission dip near $1569.7\,\mathrm{nm}$
to implement dissipative sensing. However the proposed multimode sensing
by BP-ANN is independent of the linear change of transmission depth,
and the transmission dip near $1569.7\,\mathrm{nm}$ can be still
used for multimode sensing. So the multimode sensing is robust against
the choice of transmission dip. While near $1601.5\,\mathrm{nm}$
in Fig.\ref{Fig.3}(d), the transmission depth gets smaller with the
increasing of the applied DC voltage. Near $1535.2\,\mathrm{nm}$
or $1601.5\,\mathrm{nm}$, in the voltage range of 0 to 4$\,$$\mathrm{V}$,
the transmission depth change is not strictly linear with the increase
of the applied voltage. Then the single-mode dissipative sensing could
be performed in a relatively small voltage range, within which the
dashed lines were marked in the figures.

Strictly speaking, the spectral shifts of all WGMs, including the
changes in the resonant frequency, spectrum linewidth, and transmission
depth, could be considered as the effective sensing information in
nature. Nevertheless, in WGM-based sensing fields, the conventional
approaches mainly focus on measuring the spectral change of single
resonance. The single information channel scheme wastes a lot of sensing
information from the other resonances. It is well known that the sensing
model can give more accurate results with the help of multi-sensor
fusion technology~\citep{key-31}. The sensor fusion is the ability
to bring together inputs from a set of heterogeneous or homogeneous
sensors and history values of sensor data to form a single model~\citep{key-61,key-5-18-3}.
The sensor fusion technology is based on the measurement results by
multiple sensors. The studies demonstrate that the resolution of the
target parameter by this optimal fusing process is better than a single
sensor's measurement~\citep{key-2-11-1}. In the case of SIMRR based
sensor, the spectral shifts from multiple resonances can be regarded
as multiple independent measurements by multiple sensors. The conventional
single-mode dissipative sensing can be considered as the measurement
by a single sensor. If these multimode sensing information are fused
effectively, the multimode sensing method must outperform the best
single sensor (single mode dissipative sensing).

\subsection{Cross-validation based on the artificial network}

\begin{figure*}[tp]
\centerline{\includegraphics[width=1.5\columnwidth]{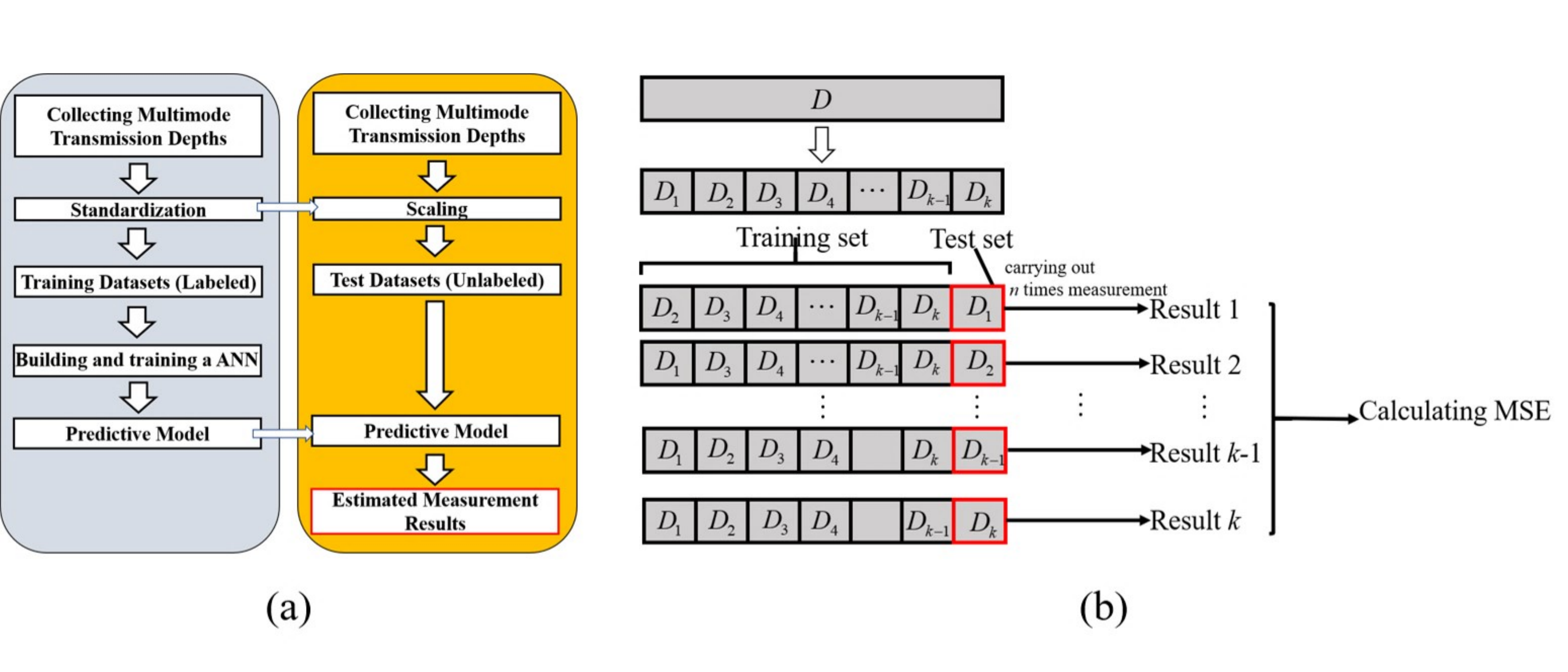}}\caption{(a) A BP-ANN based signal processing procedure for the proposed multimode
sensing. (b) Illustration of the \textsl{K}-fold cross-validation
(CV).}
\label{Fig.4}
\end{figure*}

Subsequently, the BP-ANN was successfully applied for fusing the multimode
sensing information. Fig.\ref{Fig.4}(a) displays a BP-ANN based signal
processing procedure for the proposed multimode sensing in Section~2.
In the experimental works, the BP-ANN was built to accurately estimate
the voltage applied across the microheater. First, it is necessary
to prepare the training set before the measurement process is started.
As the sensor was used in practice, the test set was obtained by collecting
a group of transmission depths within the wavelength range. It was
noted that the data with unknown output voltage (unlabeled) were independent
of the ones used for training, and the same data was not shared between
the training set and the test set. The training set was standardized.
Its data normalization could save training time and improve BP-ANN's
performance. The BP-ANN was then trained based on the training set
(labeled), resulting in the development of a prediction model. Before
the prediction model was validated with the test set, the test set
was processed with the same standardization as the training set for
scaling.

Based on the experimental results, the original data sample consists
of pairs of an input vector (a group of transmission depths) and the
corresponding setting voltage. If the original data sample is used
as a training dataset, the voltage is commonly denoted as the target
(or label). If it is selected as a test dataset, the voltage is used
to compare with the result by the BP-ANN. The supervised machine learning
models were developed through the \textit{K}-fold cross-validation
(CV)~\citep{key-19-2}. The schematic illustration of the \textit{K}-fold
CV is displayed in Fig.\ref{Fig.4}(b). In the \textit{K}-fold CV,
the original data samples are randomly partitioned into \textit{K}
subsamples. Of the \textit{K} subsamples, a single subsample is selected
as the test set for validating the model, and the remaining \textit{K-}1
subsamples are implemented as the training set. In this research,
the training and test sets were selected from the original data samples
according to the above cross-validation. Under a selection of training
and test sets, the estimate process is executed by the BP-ANN.\textcolor{black}{{}
A detailed account of the BP-ANN procedure to perform the SIMRR multimode
sensing can be found in Ref.~\citep{key-1-5}. Due to random weight
initialization in the BP-ANN before training, the output result by
the BP-ANN fluctuates when it is operated each time. Here, the estimate
process by the BP-ANN was repeatedly carried out }\textit{\textcolor{black}{n}}\textcolor{black}{{}
times, because its output accuracy could be improved by averaging
}\textit{\textcolor{black}{n}}\textcolor{black}{{} times estimate results.
To calculate the MSE of the multimode sensing, the CV process was
repeated }\textit{\textcolor{black}{K }}times (the folds), where each
of the \textit{K} subsamples is used exactly once as the test set.
We calculated the mean square errors (MSE) between \textit{K} desired
output voltages and the \textit{K} results from the folds, and the
value is considered as \textcolor{black}{the MSE of the multimode
sensing}. In the studies, there are 41 voltage values present when
the voltage applied across the micro-heater is adjusted from 0 to
4$\,$\textit{$\mathrm{V}$ }with the step size of 0.1$\,$\textit{$\mathrm{V}$.
}Then, the number of original data samples is 41, and the CV approach
is taken from 41 sub-samples (41-fold). All data samples are used
for training and testing, and each data sample is used for validation
exactly once. Such measurement process could be made without loss
of generality, and it is also an appropriate selection for objective
error estimates of the multimode sensing.

\subsection{\textcolor{black}{Results and discussion }}

\begin{figure}[p]
\centerline{\includegraphics[width=1\columnwidth]{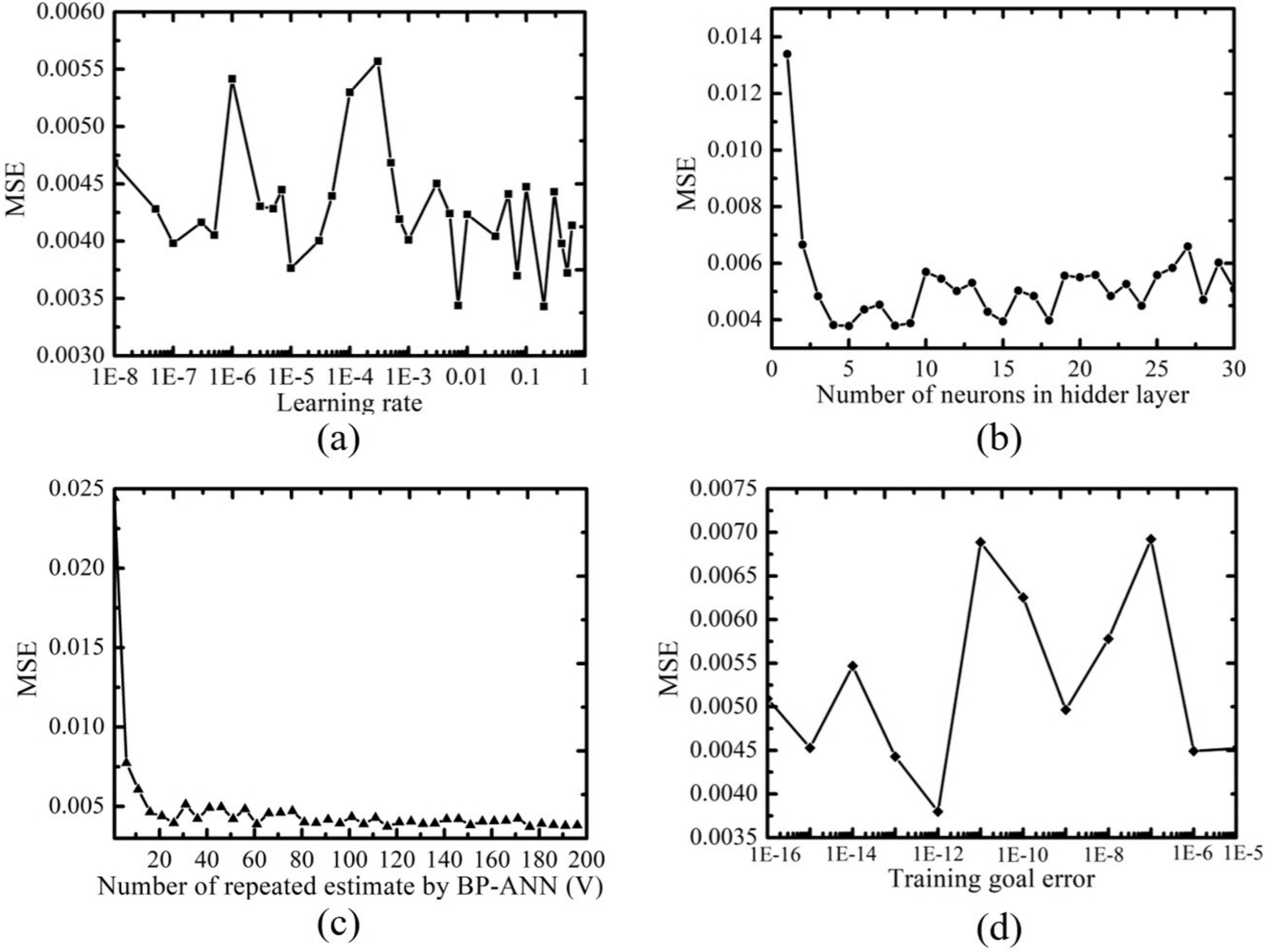}}\caption{Influence of the network parameters on its measurement results, MSE
versus (a) learning rate, (b) number of neurons in the hidden layer,
(c) number of \textcolor{black}{repeated estimate} by the BP-ANN,
and (d) training goal error.}
\label{Fig.5}
\end{figure}

\textcolor{black}{As shown in Fig.\ref{Fig.5}(a)-(b), the influence
of related parameters on measurement results, such as learning rate,
number of neurons in the hidden layer, number of repeated estimates
by the BP-ANN, and training goal error, were studied to optimize the
sensor performance. The initial parameter values were set as follows:
the node number of 10 in the hidden layer, the training goal error
of $1\times10^{-6}$, and the number of repeated estimates of 100.
Fig.\ref{Fig.5} shows the MSE versus the learning rate, and an optimal
learning rate value was selected as 0.007 with the minimum MSE. Similarly,
the number of neurons in the hidden layer, the number of repeated
measurements, and training goal error were successively optimized,
and the final parameters are selected as the node number of 5 in the
hidden layer, the training goal error of $1\times10^{-12}$, and the
number of repeated measurements of 60.}

\begin{figure}[tp]
\centerline{\includegraphics[width=1\columnwidth]{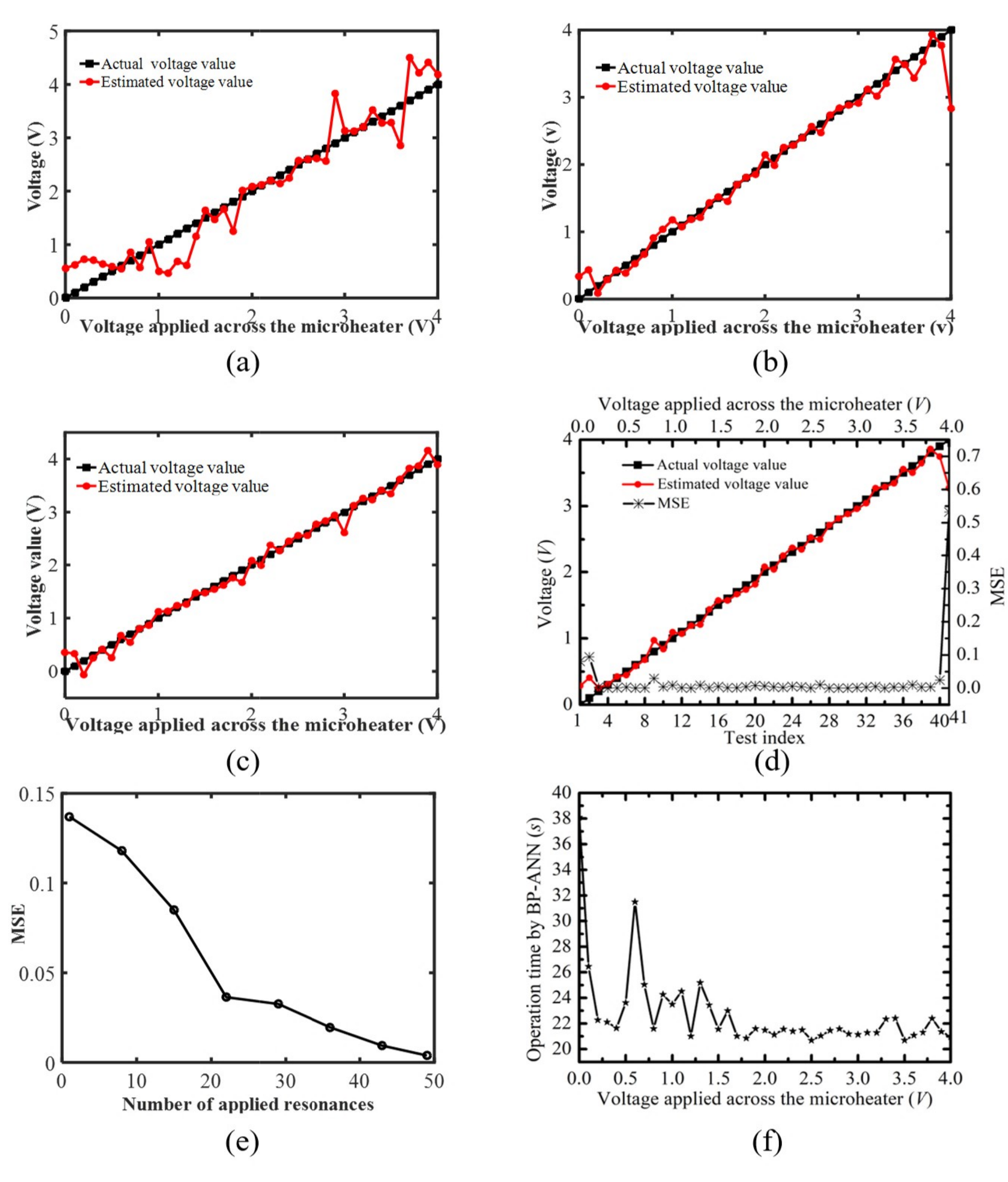}}\caption{When the numbers of applied resonances in the sensing are 1 (a), 10
(b) and 30 (c) comparison between the estimated values of the voltage
and their corresponding actual values. (d) With the number of applied
resonances \textit{$n=49$}, comparison between the estimated values
of the voltage and their corresponding actual values versus voltage
applied across the microheater, and MSE versus voltage applied across
the microheater. (e) MSE versus the number of applied resonances.
(f) Operation time by the BP-ANN versus voltage applied across the
microheater.}
\label{Fig.6}
\end{figure}

\textcolor{black}{With above-mentioned CV method, results by the BP-ANN
were compared with actual reference values} \textcolor{black}{in Fig.\ref{Fig.6}(a)-(d).
We tested our method by seeding the BP-ANN with partial experimental
data, i.e. with the number of resonances. Comparing the estimated
values of the voltage and their corresponding experimental values,
the precision increases with the number of resonances. There are few
measurement errors between measurement results and their actual experimental
values, except at the first and last two test indexes.} At several
edge test \textcolor{black}{indexes}, large measurement errors often
occur because of \textcolor{black}{the poor generalization ability
of the trained BP-ANN. To evaluate the multimode sensing's performance
fairly, four edge test indexes were removed from the voltage measurement
range. The MSE by the BP-ANN was averaged over all the test indexes
except four edge test indexes. As the numbers of applied resonances
in the proposed multimode sensing are 1, 10, 30, and 49, the values
of their MSE are 0.1377, 0.093, 0.035, and }\textit{\textcolor{black}{$3.9\times10^{-3}$}}\textcolor{black}{,
respectively, as shown in Fig.\ref{Fig.6}(e). It is proved that the
sensitivity is improved when more information inquired by the multimode
sensing method.}

\textcolor{black}{The corresponding MSE versus the voltage was also
displayed} in\textcolor{black}{{} Fig.\ref{Fig.6}$\,$(d). In the case,
as the voltage applied across the microheater was adjusted from 0.0
to 4.0}\textit{\textcolor{black}{$\,$$\mathrm{V}$}}\textcolor{black}{ 
with the step size 0.1}\textit{\textcolor{black}{$\,$}}\textcolor{black}{$\mathrm{V}$,
these original samples were numbered as test indexes from 1 to 41.
When 49 resonances in Fig.\ref{Fig.3}(a) is used in the multimode
sensing, the value of the MSE is }\textit{\textcolor{black}{$3.9\times10^{-3}$}}\textcolor{black}{.
For the multimode sensing scheme using the BP-ANN, the root-mean-square
error (RMSE) is approximately equal to its limit of detection (LOD)~\citep{key-1-5}.
Then, the LOD of the multimode sensing is }0.062$\,\mathrm{V}$. \textcolor{black}{As
a comparison, the single-mode dissipative sensing was also implemented
by measuring the change in the transmission depth of a single resonance.
The detailed derivation of its LOD was provided in Appendix A.} The
LOD of the single-mode dissipative sensing near $1535.2\,\mathrm{nm}$
and $1601.5\,\mathrm{nm}$ are $0.15\,\mathrm{V}$ and $0.293\,\mathrm{V}$,
respectively\textcolor{black}{. The LOD achieves the optimal} value
for sensing near $1535.2\,\mathrm{nm}$\textcolor{black}{, in comparison
with the LOD values at other transmission dips. The optimal LOD of
single-mode sensing is at least 2.4 times greater than that of }the
multimode sensing\textcolor{black}{. Moreover, the multimode sensing
can achieve the lower LOD within the broad voltage range of }$0.2\,\mathrm{V}\leqslant U\leqslant3.8\,\mathrm{V}$.
Therefore,\textcolor{black}{{} the experimental results proved that
the performance of the multimode sensing method is superior to that
of the single mode sensing.}

\textcolor{black}{The multimode SIMRR sensor's time resolution was
essentially set by the response time of the SIMRR, or roughly the
SIMRR photon lifetime. In the detecting system, the scanning time
by the OSA limited the time resolution of such measurements to a few
seconds. When the BP-network is operated on a personal computer with
Intel Core~i5 2.3~GHz CPU and 8~GB memory, the operation time by
the BP-ANN versus voltage applied across the microheater is displayed
in Fig.\ref{Fig.6}(f), where the measurement process by the BP-ANN
is repeatedly performed 60 times. The average operation time of the
BP-ANN is about 22.72~seconds. When the BP-ANN algorithm was run
on a computer with Intel Core~i7 3.7~GHz CPU and 16~GB memory,
the averaged operation time was shortened to 3.78~seconds. Therefore,
our approach promises the real time sensing applications with low
LOD, where the experimental data acquisition time is also on the order
of seconds. The ultimate operation time of our method is restricted
by algorithm. In practice, two approaches can be taken to employ supervised
machine learning algorithm with an even shorter time in the future.
One approach is to adopt advanced fusion algorithm (deep neural networks)
to accelerate operation and shorten the training time. The second
approach is to transmit a big multimode sensing data to the cloud
server via the 5G network, which is used to execute the algorithm
with faster speed. In such a way, it is possible to realize real-time
measurement performed on the multimode SIMRR sensor with low LOD. }

\begin{figure}[tp]
\centerline{\includegraphics[width=1\columnwidth]{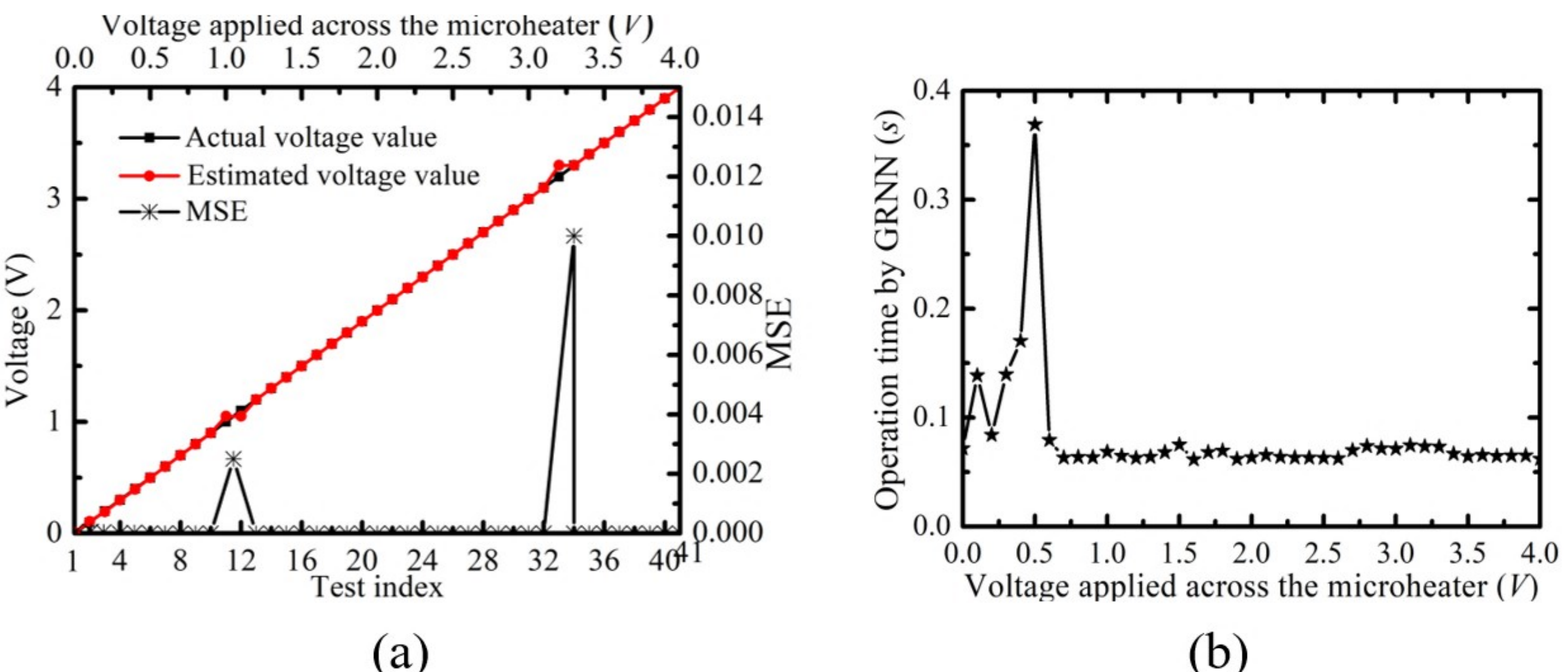}}\caption{All the intensity samples in the spectra at $1500\thinspace\mathrm{nm}\protect\leq\lambda\protect\leq1620\thinspace\mathrm{nm}$
are used to form the original data, and the GRNN algorithm is adopted
to develop the predictive model. (a) The comparison between the estimated
values of the voltage and their corresponding actual values versus
voltage applied across the microheater, and MSE versus voltage applied
across the microheater. (b) Operation time by the GRNN versus voltage
applied across the microheater.}
\label{Fig7}
\end{figure}

In the end, the BP-ANN was replaced with the general regression neural
network (GRNN) to develop the predictive model. The GRNN has an excellent
nonlinear approximation performances under small-sample data-set or
unstable data~\citep{key-19}. The GRNN consists of an input layer,
a hidden layer and an output layer where there is one hidden neuron
for each training pattern in hidden layer. The GRNN is a one-pass
learning algorithm with a highly parallel structure, and the smoothing
factor is the only free parameter in GRNN. The multimode sensing informations
are further expanded as the complete spectra signal, where the original
sensing informations can never be lost in the frequency domain. All
the intensity samples in the spectra at $1500\thinspace\mathrm{nm}\leq\lambda\leq1620\thinspace\mathrm{nm}$
were used to form the original data, including the training and test
sets. The CV method was also applied in the case. Because\textcolor{black}{ 
the output result by the }GRNN\textcolor{black}{  fluctuates slightly
when it is operated each time}. Therefore, it is unnecessary to perform
multiple estimates for each test voltage, and the GRNN was operated
only once\textcolor{black}{.} With the optimized smoothing factor
value of 1.3, the measurement outputs are shown in Fig.7 (a). It is
observed that we get the very accurate results except that there are
few errors in two test indices. The value of MSE is \textit{\textcolor{black}{$1.02\times10^{-4}$,
}}\textcolor{black}{and}\textit{\textcolor{black}{  }}\textcolor{black}{its
LOD is close to }0.01$\,\mathrm{V}$. \textcolor{black}{The optimal
LOD of single-mode sensing is almost 15 times greater than that of
}the multimode sensing\textcolor{black}{  by the GRNN. The reason
is that all the intensity samples in the transmission dips contain
complete sensing information, and their participation in }the fusion
algorithm\textcolor{black}{  significantly enhance the performance.
In addition,} Fig.\ref{Fig7}\textcolor{black}{(b) }displays the operation
time by the GRNN versus voltage applied across the microheater\textcolor{black}{.
The averaged operation time was shortened to 0.0808$\,$}\textit{\textcolor{black}{\emph{second}}}\textcolor{black}{ 
(a computer with Intel Core i7 3.7$\,$GHz CPU and 16$\,$GB memory).}

In our previous research, theoretical studies proved that the LOD
of the multimode sensing is more than two orders of magnitude lower
than that of the single mode sensing~\citep{key-1-5}. In the experiment\textcolor{black}{,
the LOD of the multimode sensing can be further improved by increasing
}the number of the training set. The estimation accuracy by the BP-ANN
partly depends on the number of the training set~\citep{key-1-5}.
In the experiments, \textcolor{black}{limited by the resolution of
the voltage (0.1$\,$$\mathrm{V}$)}, only 40 samples appear in the
training set (one-sample test) with the voltage varied from 0 to 4\textcolor{black}{$\,$$\mathrm{V}$.
Therefore, if the voltage resolution is sufficient, the training set
can be constructed with enough number of samples, and the measurement
accuracy by the ANN can exhibit a remarkable enhancement. }

\section{Conclusions}

In conclusion, a universal multimode sensing method has been experimentally
demonstrated based on the SIMRR on a silicon nitride photonic chip.
The detecting system is launched by a broad-band laser source, and
the transmission depths of multiple resonances are extracted from
the collected spectra as effective sensing features and merged by
a BP-ANN. Compared with the single-mode dissipative sensing, the proposed
SIMRR multimode sensing provides the following advantages: (i) The
BP-ANN is applied to realize the nonlinear mapping relationship between
a group of transmission depths and the target parameter to be detected.
Hence, the measurement range is no longer restricted by the linear
operation and can be expanded greatly. (ii) The multimode sensing
information is fused fully by using the BP-ANN, and the LOD can be
reduced considerably. (iii) In future application, for the detecting
system excited by a broadband laser source, only a charged coupled
device (CCD) can be used to acquire directly cavity mode intensities
in combination to a designed reflection grating. (iv) More importantly,
with abundant multimode sensing data, the multimode SIMRR sensing
method promises multi-parameter sensing applications~\citep{key-1-5}.
This will bring benefits to reduce the detection cost. The SIMRR-based
multimode sensing should thus provide a powerful platform to realize
a simple, robust, and high-sensitive sensing scheme with low detecting
cost.

\section*{Funding}

The work was supported by National Key Research and Design program
(2016YFA0301300); the National Natural Science Foundation of China
(NSFC) (60907032, 61675183 and 61675184); Zhejiang Provincial Natural
Science Foundation of China under Grant No.LY20F050009; Open Fund
of the State Key Laboratory of Advanced Optical Communication Systems
and Networks, China (2020GZKF013).

\section*{Disclosures}

The authors declare no conflicts of interest.

\setcounter{equation}{0} 
\global\long\def\theequation{A.\arabic{equation}}%

\setcounter{figure}{0} 
\global\long\def\thefigure{A.\arabic{figure}}%

\section*{Appendix A - Computation of the limit of detection of single mode
dissipative sensing in the experiment}

\begin{figure}[tp]
\centerline{\includegraphics[width=0.6\columnwidth]{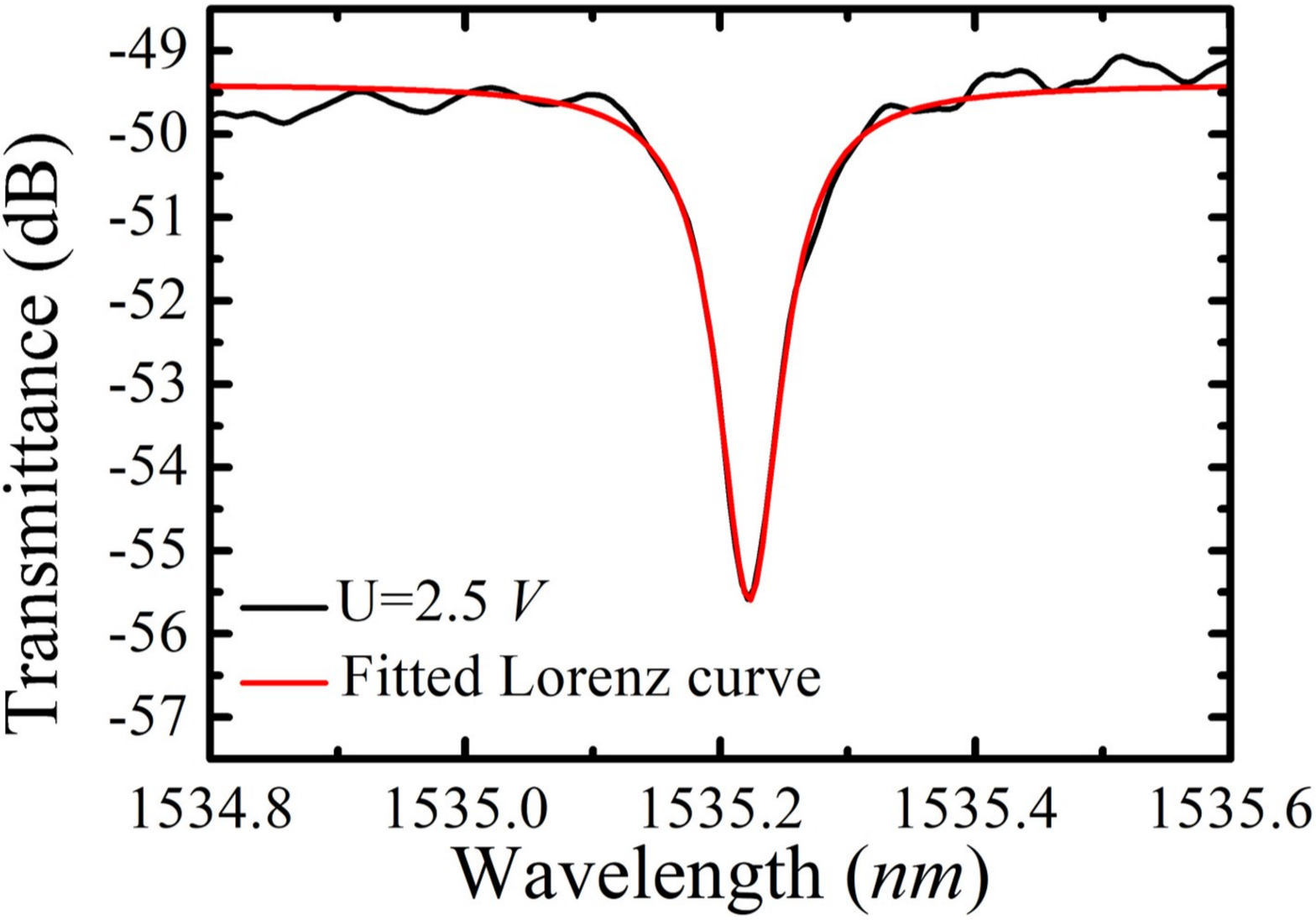}}\caption{At $U=2.5\,\mathrm{V}$, the experimental transmission spectra and
its fitted Lorentz lineshape near $1535.2\,\mathrm{nm}$, which are
indicated by the black and red solid lines, respectively.}
\label{Fig.8}
\end{figure}

The transmission depth change near $1535.2\,\mathrm{nm}$ was used
to perform the single mode dissipative sensing, and Fig.\ref{Fig.8}
shows the experimental transmission dip at \textit{$U=2.5\,\mathrm{V}$.
}They were fitted by a Lorentz lineshape curve, which can be described
by the following expression,
\begin{equation}
I=I_{0}+\frac{2A}{\pi}\frac{\Gamma}{4(\lambda-\lambda_{0})^{2}+\Gamma^{2}}
\end{equation}
where $I_{0}=-49.394\,\mathrm{dB},$$\lambda_{0}=1535.223\,\mathrm{nm},$$\Gamma=0.0594\,\mathrm{nm},$
and $A=-0.579\,\mathrm{dB}.$ The amplitude noises are the difference
between the experimental transmission spectra and its fitted Lorentz
lineshape~\citep{key-18,key-18-1}. Assuming that the amplitude noise
follows a Gaussian distribution, its standard deviation $\sigma$
was approximately equal to $0.2085$$\,\mathrm{dB}$. The LOD of the
single- mode dissipative sensing can be written as,
\begin{equation}
LOD=\frac{\sigma}{S_{I}}
\end{equation}
where $S_{I}$ denotes the single mode dissipative sensitivity. Fig.\ref{Fig.3}(b)
shows the experimental transmission dip near $1535.2\,\mathrm{nm}$
as the applied voltage was changed. The single-mode dissipative sensing
was realized by measuring the transmission depth changes with different
sensitivities ($S_{I}$). Aided by the dashed line marked in Fig.\ref{Fig.3}(b),
the averaged sensitivity is about $1.389\,\mathrm{dB/V}$ at $1.7\,\mathrm{V}\leqslant U\leqslant3.5\,\mathrm{V}$.
Hence, the LOD of the single-mode dissipative sensing is 0.15$\,\mathrm{V}$
according to Eq.$\,$(2). In the same way, for the transmission dips
in Fig.3$\,$(d), its LOD value is about $0.293\,\mathrm{V}$, where
the noise standard deviation $\sigma$ is $0.3779$$\,\mathrm{dB}$,
and its average sensitivity is 1.293$\,\mathrm{dB/V}$ at $1.3\,\mathrm{V}\leqslant U\leqslant3.3\,\mathrm{V}$.

\end{document}